\documentclass[aps,preprint]{revtex4}%
\usepackage{amsfonts}
\usepackage{amsmath}
\usepackage{amssymb}
\usepackage{graphicx}%
\providecommand{\U}[1]{\protect \rule{.1in}{.1in}}

\begin{document}
\preprint{ }
\title[Sturmians]{Iterative solution of integral equations on a basis of positive energy
Sturmian eigenfunctions. }
\author{George Rawitscher}
\affiliation{Department of Physics, University of Connecticut, Storrs, CT 06269}
\pacs{PACS number}

\begin{abstract}
Years ago S. Weinberg suggested the "Quasi-Particle" method (Q-P) for
iteratively solving an integral equation, based on an expansion in terms of
sturmian functions that are eigenfunctions of the integral kernel. An
improvement of this method is presented that does not require knowledge of
such sturmian functions, but uses simpler auxiliary sturmian functions
instead. This improved Q-P method solves\ the integral equation iterated to
second order so as to accelerate the convergence of the iterations. Numerical
examples are given for the solution of the Lippmann-Schwinger integral
equation for the scattering of a particle from a potential with a repulsive
core. An accuracy of $1:10^{6}$ is achieved after $14$ iterations, and
$1:10^{10}$ after $20$ iterations. The calculations are carried out in
configuration space for positive energies with an accuracy of $1:10^{11}$ by
using a spectral expansion method in terms of Chebyshev polynomials. The
method can be extended to general integral kernels, and also to solving a
Schr\"{o}dinger equation with Coulomb or non-local potentials.

\end{abstract}
\startpage{1}
\endpage{ }
\maketitle

\section{Introduction}

Negative energy sturmian functions (S-F), introduced in the 60'ties by M.
Rotenberg \cite{ROTENBERG}, found many useful applications, such as in the
calculation of electron induced ionization collisions \cite{IONIZ2},
\cite{IONIZ}, in the identification of resonances in nucleon-nucleus
scattering \cite{CANTON}, in the solution of a Schr\"{o}dinger equation with
non-local potentials \cite{NONLOC}, for a separable representation of
scattering $t-$matrices \cite{T-MATRIX}\ and in the solution of three-body
Faddeev equations \cite{SANDHAS}. For the applications involving long range
Coulomb forces, the analytical expressions for the Coulomb Green's functions,
initially developed by L. Hosteler and R. Pratt \cite{HOSTELER} have found
many applications \cite{MAQUET}. sturmian functions form a complete, discrete
set of eigensolutions of a Sturm-Liouville differential (or integral)
equation, and form a complete set of basis functions. However the expansion of
a wave function into a set of sturmian functions in many cases does not
converge well \cite{CONV}, and methods to improve the convergence, such as
Pad\'{e} approximations, have frequently been utilized \cite{PADE},
\cite{IONIZ2}. The atom-atom potentials that occur in atomic physics
calculation often have strong repulsive cores, in addition to a long range
attractive part, and for such potentials it has been difficult to obtain a
reliable set of sturmian functions. A method based on spectral expansions in
terms of Chebyshev functions has been recently developed \cite{I-STRING} that
overcomes this difficulty, and hence the use of sturmian functions is again of
interest. Nevertheless, for numerical calculations the expansions have to be
truncated, and because of the slow convergence of sturmian expansions, an
iterative method to correct for the truncation error becomes desirable. It is
the purpose of this study to introduce such a method, based on the original
Quasi-Particle method of Weinberg \cite{WEINBERG}, and, for the purpose of
testing it, apply it to the scattering of a particle from a potential as
described by the Lippmann-Schwinger (L-S) integral equation. Since, the method
described here should be applicable to more general kernels of an integral
equation, the present study is done in anticipation of solving the more
complicated two-dimensional integral equations that occur in the solution of
three-body equations in configuration space \cite{CONV}, \cite{THREE-R}. The
expected accuracy is to be better than 6 to 7 significant figures, desirable
for doing atomic physics calculations. By comparison, the solution of
three-body equations for nuclear physics applications, done commonly in
momentum space \cite{THREE-P}, achieve an accuracy not better than four
significant figures \cite{GLOECKLE}.

The present paper examines the convergence of several iteration methods for
the solution of the L-S equation describing the scattering of a particle with
positive energy $E$ by a local potential\ $V$ in one dimension. The insights
gained from this example is to guide one for solving a more general integral
equation. The method initially suggested by Weinberg \cite{WEINBERG}, also
called the "Quasi-Particle" method Q-P, requires sturmian functions that are
the eigenstates for the exact integral operator. The present method, denoted
as method $\mathcal{S}_{2},$\  \ differs from the Q-P method in that it uses
auxiliary sturmian functions that are based on an auxiliary potential $\bar
{V}$ that is not the same as the potential $V.$ Since it may be
computationally easier to obtain the auxiliary sturmians rather than the
sturmians for the original integral operator method $\mathcal{S}_{2}$ may
prove advantageous for the case of very complicated integral kernels. For the
$L-S$ example, the auxiliary sturmians are used to approximate the scattering
potential $V(r)$ by a non-local separable representation $V_{N}(r,r^{\prime})$
in terms of $N$ auxiliary sturmian functions and subsequently correcting for
the truncation error $V_{R}=V-V_{N}$ by means of iterations. We find that the
rate of convergence of the iterations does not depend significantly on the
choice of $\bar{V}$, provided that the range of $\bar{V}$ is sufficiently
larger than the range of $V$. The sturmian functions in all these cases are
calculated in configuration space for a positive energy $E$, so that the
asymptotic form of the approximated wave function is the same as the exact
wave function. When solving for a bound-state wave function, it is customary
the represent the potential by negative energy sturmians that are real, but
for scattering cases, positive energy sturmians should be preferable. An
advantage of the sturmian expansion method over the Fourier-Grid method
\cite{KOSLOFF} for positive energies is that the sturmian method emphasizes
only the spatial region where the potential is non-negligible, the asymptotic
part of the wave function being already incorporated into the sturmian basis,
while in a Fourier-Grid method, the asymptotic part has to be obtained explicitly.

Since an accuracy of the scattering solution to $1:10^{8}$ is envisaged, the
sturmian functions are also required to have the same, or better, accuracy.
Our approach for solving the Sturm-Liouville eigenvalue equation
\cite{I-STRING} utilizes a "spectral" expansion method into Chebyshev
polynomials \cite{SPECTRAL-A}, \cite{SPECTRAL-B}, that permits one to
prescribe a desired accuracy. With this method one can obtain sturmian
eigenvalues and eigenfunctions with an accuracy of, for example, $1:10^{11},$
and because of its stability, one can incorporate the effect of long-range
tails of potentials, as was demonstrated in the calculation of the bound state
eigenvalue of a Helium-Helium dimer \cite{HEHE}. The present calculations are
performed in configuration space, and can be generalized to include asymptotic
Coulomb functions. Contrary to what is often done at negative energies, our
approach does not expand the Green's functions into a separable set of
sturmians, but rather expands the potential $V(r)$ into such a set.
The\ present method to calculate sturmian functions is similar to a method
introduced previously \cite{RAW}, in that it does not use a square well
potential sturmian basis set in terms of which the desired sturmians are
expanded, and, being based on a spectral method, is considerably more precise.
This additional feature permits a more accurate study of the iteration
convergence properties.

In section II we present the formalism that defines the sturmian functions;
section III describes the approximate Q-P methods $\mathcal{S}_{1}$ and
$\mathcal{S}_{2}$ that extend Weinberg's Q-P original iterative method so that
the use of sturmian functions for the integral kernel are avoided, and
auxiliary sturmians are used instead; in section IV numerical examples are
presented for the case that the integral kernel is of the Lippmann-Schwinger
type, and section V contains the summary and conclusions. Appendices A, B, C
and D contain further details.

\section{Notation and Equations}

The one-dimensional Schr\"{o}dinger equation for the $L=0$ radial wave
function $\psi(r)$ is%
\begin{equation}
\mathcal{D}\  \psi=V\psi \label{1}%
\end{equation}
where
\begin{equation}
\mathcal{D=}\frac{d^{2}}{dr^{2}}+E \label{2}%
\end{equation}
is the differential operator, $E$ is the energy assumed positive here, $V$ is
the scattering potential. These quantities are in units of inverse length
squared, and were transformed from their energy units by multiplication with
the well known factor $2\mu/\hbar^{2}.$ The wave number $k$ is related to the
energy by $E=k^{2}.$ The corresponding Lippmann-Schwinger integral equation
for $\psi$ is%
\begin{equation}
\psi(r)=F(r)+\int_{0}^{\infty}\mathcal{G}_{0}(r,r^{\prime})V(r^{\prime}%
)\psi(r^{\prime})dr^{\prime}. \label{3}%
\end{equation}
The undistorted Green's function is given by
\begin{equation}
\mathcal{G}_{0}(r,r^{\prime})=-\frac{1}{k}F(r_{<})\times H(r_{>})\text{ }
\label{4}%
\end{equation}
where $(r,r\prime)=(r_{<},r_{>})$ if $r\leq r^{\prime}$ and $(r,r\prime
)=(r_{>},r_{<})$ if $r\geq r^{\prime}$. In the present zero angular momentum
case%
\begin{equation}
F(r)=\sin(kr);~~~H(r)=\cos(kr)+i\sin(kr). \label{5}%
\end{equation}
According to Eqs. (\ref{3}) and (\ref{4}) the asymptotic form of $\psi$ is%
\begin{equation}
\psi(r\rightarrow \infty)=F(r)+S\ H(r) \label{6}%
\end{equation}
with
\begin{equation}
S=-\frac{1}{k}\int_{0}^{\infty}F(r^{\prime})V(r^{\prime})\psi(r^{\prime
})\ dr^{\prime}. \label{7}%
\end{equation}
Near the origin $\psi(r\rightarrow0)=0,$ because both $F$ as well as the
integral term in Eq. (\ref{3}) go to zero.

The sturmian functions $\Phi_{s}(r)$, $s=1,2,...$ obey Eq. (\ref{1}) with $V$
replaced by $\Lambda \  \bar{V}$%
\begin{equation}
\mathcal{D}\  \Phi_{s}=\Lambda_{s}\bar{V}\  \Phi_{s}, \label{8}%
\end{equation}
where $\bar{V}(r)$ is the sturmian potential chosen conveniently, that need
not be equal to the scattering potential $V(r)$, and the $\Lambda$'s are the
eigenvalues to be determined. The sturmian functions \ $\Phi_{s}$ obey the
boundary conditions%
\begin{equation}
\Phi_{s}(r\rightarrow0)=0;~~\Phi_{s}(r\rightarrow \infty)=\mathfrak{k}%
_{s}\mathfrak{\ }H(r), \label{9}%
\end{equation}
where the constant $\mathfrak{k}_{s}$ is determined by the normalization of
the sturmian function. The functions $\Phi_{s}$ obey the integral equation
version of Eq. (\ref{8})
\begin{equation}
\eta_{s}\Phi_{s}(r)=\int_{0}^{\infty}\mathcal{G}_{0}(r,r^{\prime})\bar
{V}(r^{\prime})\Phi_{s}(r^{\prime})dr^{\prime},\  \  \ s=1,2,3,.... \label{10}%
\end{equation}
with the Green's function given by Eq. (\ref{4}) for positive energies. The
$\eta$'s are eigenvalues of the operator $\mathcal{G}_{0}\bar{V}$, they form
an infinite set with point of convergence at $0$, and are related to the
$\Lambda$'s according to $\eta_{s}=1/\Lambda_{s}.$ With the choice (\ref{4})
and (\ref{5}) of the Green's function the sturmians $\Phi_{s}$ obey the
boundary conditions (\ref{9}), and
\begin{equation}
\mathfrak{k}_{s}=-\frac{1}{k\  \eta_{s}}\int_{0}^{\infty}F(r)\bar{V}(r)\Phi
_{s}(r)dr \label{10a}%
\end{equation}

The $\Phi$'s are not square integrable for positive energies or orthogonal to
each other. However, they are orthogonal if one includes the potential
$\bar{V}$ as a weight function%
\begin{equation}
<\Phi_{s}|\bar{V}\  \Phi_{s^{\prime}}>\  \equiv \int_{0}^{\infty}\  \Phi
_{s}(r)\  \bar{V}\  \Phi_{s^{\prime}}(r)dr=0\text{ for }\eta_{s}\neq
\eta_{s^{\prime}}. \label{11}%
\end{equation}
In the bra-ket notation above, $<\Phi_{s}|$ is \emph{not} the complex
conjugate of $\Phi_{s}$, as is usually implied. The validity of (\ref{11}) can
be shown by replacing $\Phi_{s^{\prime}}\rangle$ by $\Lambda_{s^{\prime}%
}\mathcal{G}_{0}\bar{V}\Phi_{s^{\prime}}\rangle$ in the integral (\ref{11}),
and subsequently replacing $\langle \Phi_{s}\bar{V}\  \mathcal{G}_{0}$ by
$\eta_{s}\langle \Phi_{s}$. The result is $<\Phi_{s}|\bar{V}\  \Phi_{s^{\prime}%
}>\ =\  \eta_{s}\Lambda_{s^{\prime}}<\Phi_{s}|\bar{V}\  \Phi_{s^{\prime}}>$, and
if $\eta_{s}\Lambda_{s^{\prime}}\neq1$, the above identity becomes absurd
unless $<\Phi_{s}|\bar{V}\  \Phi_{s^{\prime}}>\ =0.$ Because of the
completeness of the sturmian functions, one has the identity%
\begin{equation}
\delta(r-r^{\prime})=\sum_{s=1}^{\infty}\Phi_{s}(r)\frac{1}{\langle \Phi
_{s}\bar{V}\Phi_{s}\rangle}\Phi_{s}(r^{\prime})\bar{V}(r^{\prime}), \label{A6}%
\end{equation}
which can also be written as
\begin{equation}
\delta(r-r^{\prime})=\sum_{s=1}^{\infty}\bar{V}(r)\Phi_{s}(r)\frac{1}%
{\langle \Phi_{s}\bar{V}\Phi_{s}\rangle}\Phi_{s}(r^{\prime}), \label{B6}%
\end{equation}
depending on whether the delta function applies to the left or to the right of
the integrand in an integral.

One can understand intuitively the properties of the $\Phi$'s as follows
\cite{RAW}. As the index $s$ increases, the corresponding values of
$\Lambda_{s}$ increase, and hence the potential $\Lambda_{s}\bar{V}$ increases
in magnitude. If $\bar{V}$ is real and attractive and the real part of
$\Lambda_{s}$ is positive, then the real part of $\Lambda_{s}\bar{V}$ becomes
more attractive, and the corresponding eigenfunction $\Phi_{s}$ becomes more
oscillatory inside of the attractive region of the well. So, from one $s$ to
the subsequent one $s+1$ the eigenfunction acquires one more node inside of
the well. According to flux considerations the imaginary part of $\Lambda
_{s}\bar{V}$ has to be positive, i.e., the well has to be emissive \cite{RAW}.
Near the origin the flux is $\simeq0$ since $\Phi \simeq0$ however
asymptotically the outgoing form of the wave function produces a positive
outgoing flux. This outgoing flux is generated by an emissive imaginary
potential, exactly the opposite of\ the case of an optical potential, that
absorbs flux. These properties will be verified in the numerical section below.

\subsection{Numerical properties of sturmian functions.}

\begin{figure}
[ptb]
\begin{center}
\includegraphics[
height=1.7538in,
width=2.3324in
]%
{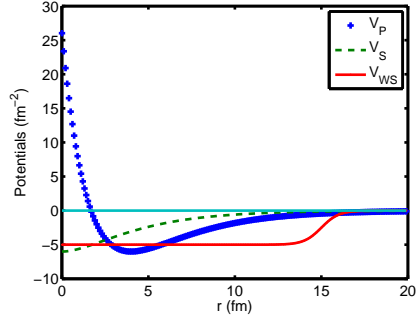}%
\caption{Three sturmian potentials as a function of radial distance, given by
Eqs. (\ref{21}) and (\ref{21WS}).}%
\label{FIG1}%
\end{center}
\end{figure}

In the examples given below, the scattering potential $V$ is of the Morse type
with a repulsive core near the origin ($V_{P}$),\ and the sturmian potential
$\bar{V}$\ is either also of the Morse type with the repulsive core suppressed
($V_{S})$, but of the same range as the scattering potential $V_{P}$, or it is
a potential of the Woods-Saxon type ($V_{WS}$) with a range larger than that
of either $V_{P}$ or $V_{S}.$ The Morse potentials are formed by the
combination of two exponentials%
\begin{equation}
\bar{V}(r)=V_{M}\  \exp(-r/\alpha+\beta/\alpha)\times \left[  \exp
(-r/\alpha+\beta/\alpha)-2\right]  , \label{21}%
\end{equation}
with parameters given in Table \ref{TABLE1}
\begin{table}[tbp] \centering
\begin{tabular}
[c]{|l||l|l|l|}\hline
& $V_{M}\ (fm^{-2})$ & $\  \  \alpha \ (fm)$ & $\beta \ (fm)$\\ \hline \hline
$V_{P}$ & $6$ & $1/0.3$ & $4$\\ \hline
$V_{S}$ & $6$ & $1/0.3$ & $0$\\ \hline
\end{tabular}
\caption{Parameters for two Morse-type potentials P and S.}\label{TABLE1}%
\end{table}%
, and the Woods-Saxon potential is given by%
\begin{equation}
V_{WS}=V_{0}/\{1-\exp[(r-R)/a]\} \label{21WS}%
\end{equation}
with $V_{0}=-5\ fm^{-2},\ R=15\ fm$ and $a=0.5\ fm.$The resulting dependence
of these potentials on $r$ is illustrated in Fig. \ref{FIG1}.%

\begin{figure}
[ptb]
\begin{center}
\includegraphics[
height=1.8265in,
width=2.4267in
]%
{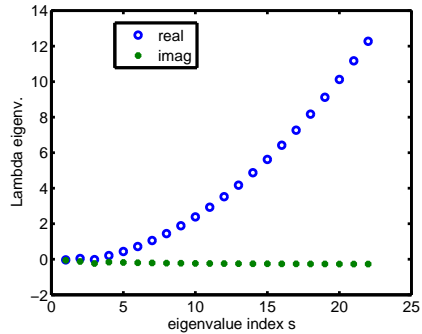}%
\caption{The spectrum of the sturmian eigen values $\Lambda$ for the Morse
Potential $V_{S}$ defined in Table \ref{TABLE1} and illustrated in Fig.
\ref{FIG1}. The wave number is $k=0.5\ fm^{-1}.$}%
\label{FIG2}%
\end{center}
\end{figure}

\begin{figure}
[ptb]
\begin{center}
\includegraphics[
height=1.8887in,
width=2.5105in
]%
{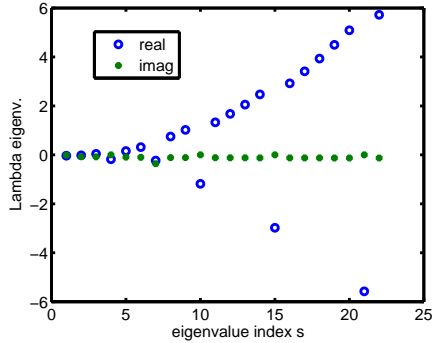}%
\caption{Same as Fig. \ref{FIG2} for the potential $V_{P}$ \ defined in Table
\ref{TABLE1}. This potential has a repulsive core.}%
\label{FIG3}%
\end{center}
\end{figure}

\begin{figure}
[ptb]
\begin{center}
\includegraphics[
height=1.8369in,
width=2.4422in
]%
{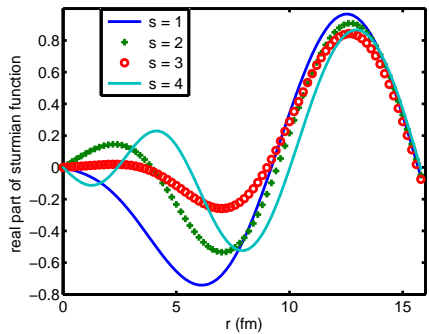}%
\caption{Sturmian eigenfunctions for the potential $V_{S}$ defined in Table
\ref{TABLE1} for a wave number $k=0.5\ fm^{-1}.$ They are normalized such that
asymptotically they all approach the outgoing Hankel function
$H(r)=cos(kr)+i\ sin(kr).$}%
\label{FIG4}%
\end{center}
\end{figure}

\begin{figure}
[ptb]
\begin{center}
\includegraphics[
height=1.8421in,
width=2.4483in
]%
{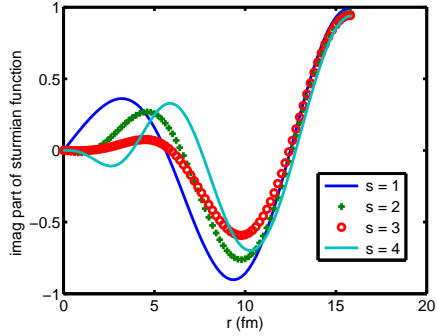}%
\caption{Same as Fig \ref{FIG4} for the imaginary part of the sturmian
functions.}%
\label{FIG5}%
\end{center}
\end{figure}

The spectrum of the eigenvalues $\Lambda_{s}$ for the two Morse-type
potentials, with $k=0.5fm^{-1}$ is illustrated in Figs. \ref{FIG2} and
\ref{FIG3}.
Since the potential $V_{S}$ is entirely attractive, the real parts of
$\Lambda_{s}$ are positive, while the imaginary parts are negative, in
accordance with the argument given above. These sturmian eigenfunctions are
obtained by an adaptation of Hartree's iterative method designed to obtain
energy eigenvalues to the Schr\"{o}dinger equation, and described in Appendix
C. A list of these eigenvalues precise to $11$ significant figures is given in
the Appendix D. Some of the eigenfunctions are illustrated in Figs. \ref{FIG4}
and \ref{FIG5}, that show that at large distances\ where the potential becomes
small, these functions become linearly dependent. The $\Phi_{s}$ used in the
present examples are normalized such that asymptotically they equal the
function $H(r)$, with unit coefficient.

\begin{figure}
[ptb]
\begin{center}
\includegraphics[
height=1.8939in,
width=2.5175in
]%
{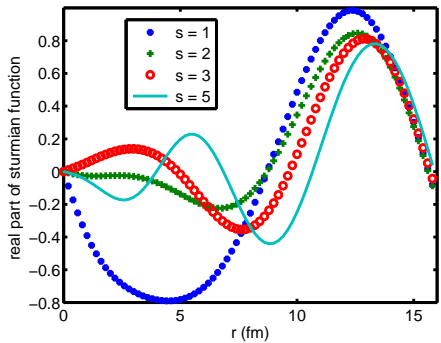}%
\caption{Real parts of sturmian functions $\Phi_{s}$ for the potential $V_{P}%
$, for $k=0.5\ fm^{-1}$. This potential, defined in Table \ref{TABLE1}, has a
repulsive core.}%
\label{FIG6}%
\end{center}
\end{figure}

\begin{figure}
[ptb]
\begin{center}
\includegraphics[
height=1.9934in,
width=2.6489in
]%
{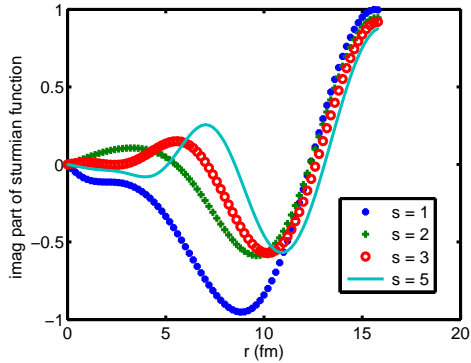}%
\caption{Same as Fig. \ref{FIG6} for the imaginary parts of the sturmian
functions}%
\label{FIG7}%
\end{center}
\end{figure}

\begin{figure}
[ptb]
\begin{center}
\includegraphics[
height=1.8317in,
width=2.4344in
]%
{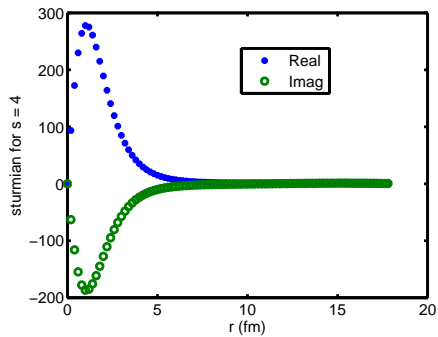}%
\caption{Sturmian function in category II for $s=4,$ for potential $V_{P}$ and
$k=0.5.$}%
\label{FIG8}%
\end{center}
\end{figure}

\begin{figure}
[ptb]
\begin{center}
\includegraphics[
height=1.9458in,
width=2.5867in
]%
{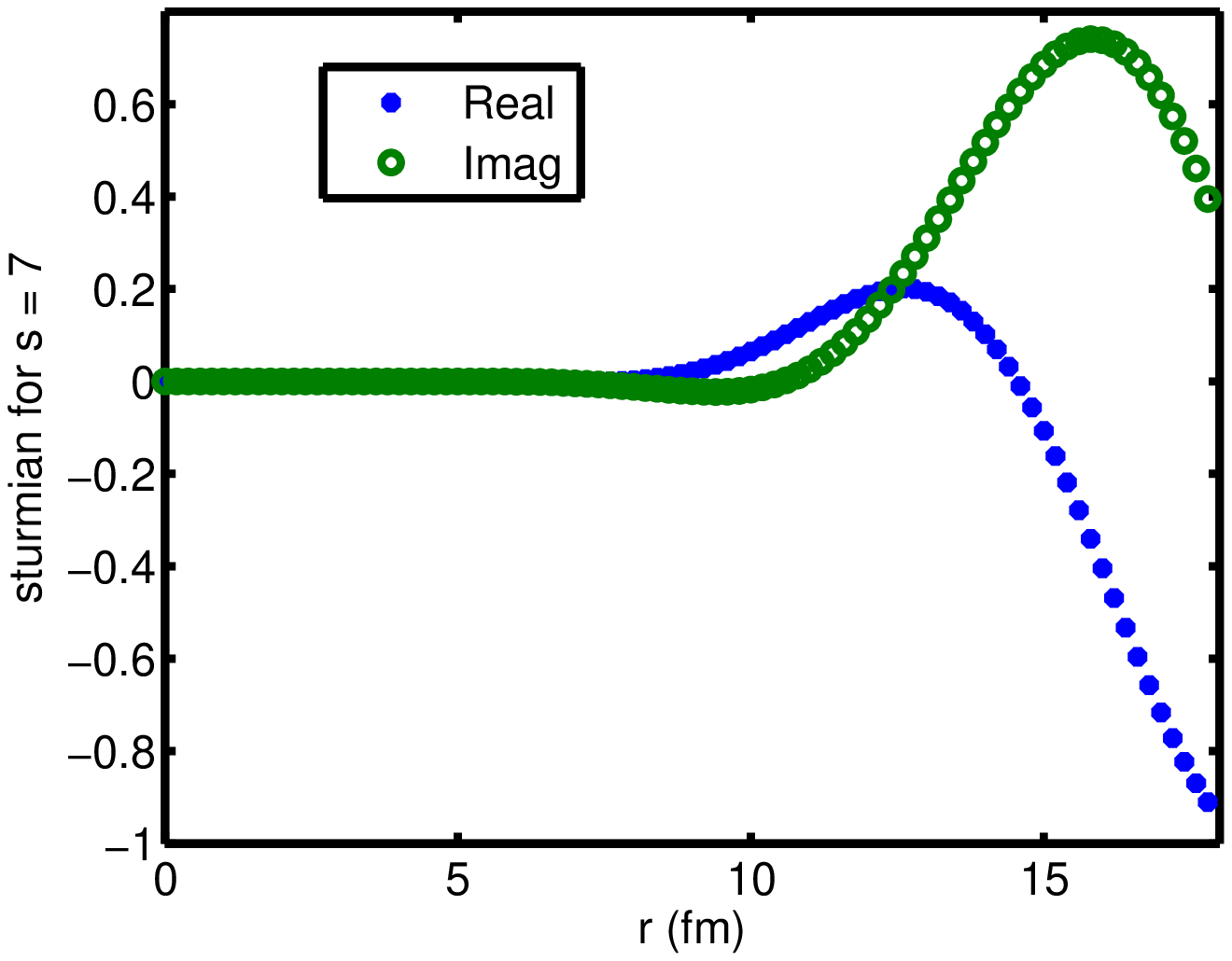}%
\caption{Same as for Fig. \ref{FIG8}, for $s=7$}%
\label{FIG9}%
\end{center}
\end{figure}

\begin{figure}
[ptb]
\begin{center}
\includegraphics[
height=1.9104in,
width=2.5408in
]%
{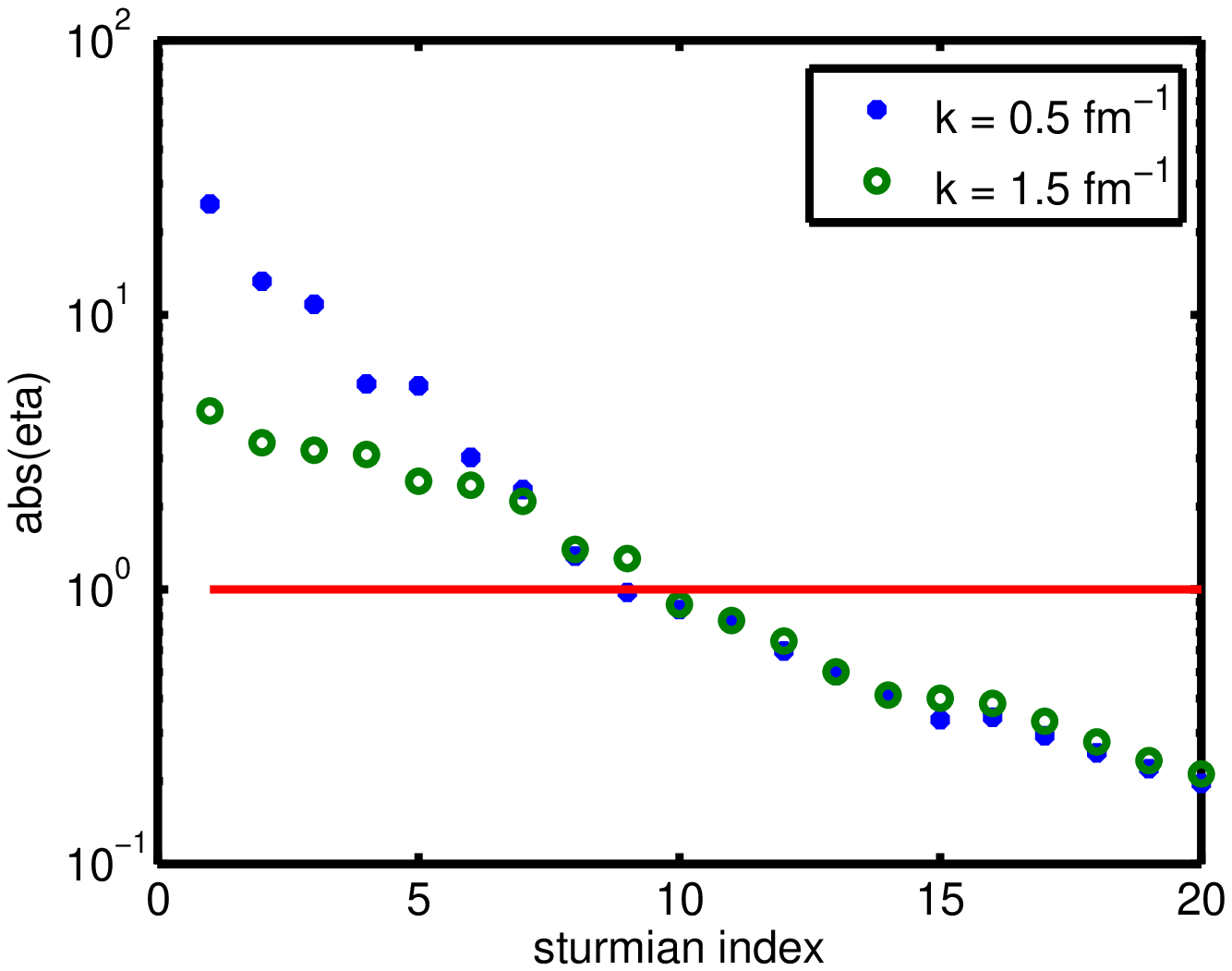}%
\caption{Absolute value of $\eta_{s}$ as a function of the sturmian index $s$
for two values of the wave number $k.$ The eigenvalues $\eta_{s}$ are defined
in Eq. (\ref{10}) with $\bar{V}=V_{P}.$}%
\label{FIG11}%
\end{center}
\end{figure}

\begin{figure}
[ptb]
\begin{center}
\includegraphics[
height=1.8317in,
width=2.4344in
]%
{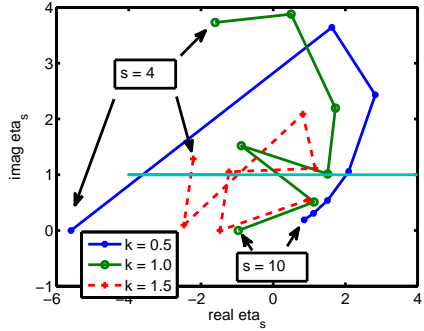}%
\caption{Argand diagram of $\eta_{s}$, for the potential $V_{P}$ for
$s=4,5,..10,$ for three values of $k$ (in units of $fm^{-1}).$ For $s=7$ they
lie close to unity.}%
\label{FIG14}%
\end{center}
\end{figure}

Because potential $V_{P}$ has both a repulsive and an attractive part, the
eigenvalues fall into two categories. In category I the eigenvalues $\Lambda$
have a positive real part and a negative imaginary part, and the corresponding
eigenfunctions are large mainly in the attractive regions of the potential
well. Examples are given in Figs. \ref{FIG6} and \ref{FIG7}.%
In category II the real parts of $\Lambda$ are negative so as to turn the
repulsive piece of the potential near the origin into an attractive well, and
the formerly attractive valley into a repulsive barrier. Examples of the
corresponding sturmian for indices $s=4,7,10,15,$ and two of these functions
are shown in Figs. \ref{FIG8} and \ref{FIG9}.
For $0<r<4\ fm$ the function $\Phi_{7}$ has a magnitude less than $10^{-4}.$
The sturmian for $s=10$ is similar to that for $s=4$ in that it is also large
near the origin (with an amplitude of $\simeq10^{9})$ and has a node near
$r=1.$ The functions for $s=4$ and $10$ are "resonant" in the radial region
$r\in \lbrack0,4]$, while the one for $s=7$ is non-resonant. For a larger
energy the effect of the barrier for the functions of class II decreases, and
the absolute value of the eigenvalues $\eta_{s}=1/\Lambda_{s}$ decreases for
$s<10,$ as is illustrated if Fig. \ref{FIG11}.
As will be shown in section III the functions for which $\eta_{s}<1$ play an
important role in the Q-P method for the iterative correction of the
truncation errors. Further, in the expansion of a wave function in terms of
sturmians, such as described in section IV-A below, a dominator $(1-\eta_{s})$
is likely to appear. For the values of $s$ for which $\operatorname{real}%
(\eta_{s})\simeq1$, and $imag(\eta_{s})\simeq0$, the corresponding sturmians
make a resonant contribution to the expansion. In Fig. \ref{FIG14} some of the
$\eta_{s}$ are illustrated in the form of an Argand diagram for three values
of $k,$ which shows that for $s=7$ the values of $\eta_{7}$ for all three
values of $k$ satisfy this resonance criterion by lying close to unity.

\section{Approximate Quasi Particle methods $\mathcal{S}_{1}$ and
$\mathcal{S}_{2}$}

The original Q-P iteration procedure was developed for the solution of the
$L-S$ Eq. (\ref{3}), but is couched here in terms of a more general integral
kernel $\mathcal{O}$ for a one dimensional integral equation
\begin{equation}
\psi(r)=F(r)+\int_{0}^{\infty}\mathcal{O(}r,r^{\prime})\psi(r^{\prime
})\ dr^{\prime}. \label{22}%
\end{equation}
This equation is to be solved for $\psi$, and $F$ is a known "driving" term.
The shorthand form of Eq. (\ref{22}) is%
\begin{equation}
\psi=F+\mathcal{O}\psi.\  \label{22a}%
\end{equation}
One procedure, convenient if $\mathcal{O}$ has a semi-separable nature, as is
the case for Green's functions in configuration space, is to expand the
unknown function into a set of Chebyshev polynomials, as is done in the IEM
method \cite{SPECTRAL-A}, and obtain an answer\ to high accuracy without
iterations. If however the operator is very complex and of large norm, so that
a Born series would not converge, then an alternative method based on
expansions into sturmian functions, described by Weinberg \cite{WEINBERG} as
the "Quasi-Particle" method, is as follows. One approximates the operator
$\mathcal{O}$ by a separable part of rank $N$, $\mathcal{O}_{N},$and the
remainder $\mathcal{O}_{R}$ is defined as%
\begin{equation}
\mathcal{O}_{R}=\mathcal{O}-\mathcal{O}_{N}, \label{27}%
\end{equation}
and iterates on the remainder. If the norm of $\mathcal{O}_{R}$ is less than
unity, the iterations should converge. Since the numerical complexity of
performing iterations is less than solving a linear equation with a matrix of
large dimension, Weinberg's method can be computationally
advantageous.\ Weinberg's method to obtain \ $\mathcal{O}_{N}$ is to calculate
the sturmians $\Gamma_{i}$ associated with the operator $\mathcal{O}$ which
satisfy the same boundary conditions as $\psi$
\begin{equation}
\mathcal{O}\Gamma_{i}=\gamma_{i}\Gamma_{i}. \label{24}%
\end{equation}
and approximate $\mathcal{O}$ by a separable expansion into those sturmians
$\Gamma_{i}$ whose eigenvalues $\gamma_{i}$, $i=1,2,..,N$\ have a magnitude
larger than unity. Iterations on the remainder will then converge, since they
involve powers of $\gamma_{i}$ which for $i>N$ are less than unity. A
numerical example (example $P$) will be given for the case that $\mathcal{O=G}%
V,$ where \ the potential $V=V_{P}$ of Eq. (\ref{21})\ has a repulsive core
and an attractive well.

Since the computational effort to obtain the eigenvectors and eigenvalues of a
general operator $\mathcal{O}$ can be larger than the effort to solve Eq.
(\ref{22}) directly, a simplified method (method $\mathcal{S}_{1}$
$\mathcal{)}$ will be described below, that uses auxiliary sturmian functions.
These methods avoid the rigorous calculation of the $\Gamma_{i}$ by expanding
the operator $\mathcal{O}$ into a finite set of auxiliary sturmian functions
$\Phi_{s}$, $s=1,2,...,N,$ which satisfy Eq. (\ref{10}) where $\bar{V}$ is an
auxiliary potential,\ and which obey the same boundary conditions as the
function $\psi.$%
\begin{equation}
\mathcal{O}_{N}\  \mathcal{=}\sum_{s=1}^{N}\mathcal{O\ }\Phi_{s}\rangle \frac
{1}{\langle \Phi_{s}\bar{V}\Phi_{s}\rangle}\langle \Phi_{s}\bar{V}.\label{25}%
\end{equation}
These sturmian functions can be obtained reliably and with high accuracy,
using a spectral expansion method \cite{SPECTRAL-B}, \cite{I-STRING}.However,
the auxiliary sturmians are not used to obtain the eigenfunctions for the
operator $\mathcal{O},$ as is done in other studies \cite{CANTON} where the
emphasis was in obtaining the scattering resonances, but rather to set up a
well converging iteration scheme. In method $\mathcal{S}_{1}$ , designed to
solve Eq. (\ref{22}) iteratively one first defines a function $\mathcal{F}%
^{(N)}$\ which is the solution of
\begin{equation}
\mathcal{F}_{1}^{(N)}=F+\mathcal{O}_{N}\mathcal{F}_{1}^{(N)}.\label{30}%
\end{equation}
Since $\mathcal{O}_{N}\ $is given by the separable representation (\ref{25}),
a solution of Eq. (\ref{30}) is of the form
\begin{equation}
\mathcal{F}_{1}^{(N)}=F+\sum_{s^{\prime}=1}^{N}\  \mathcal{O\ }\Phi_{s}%
\rangle \ c_{s^{\prime}}.\label{42}%
\end{equation}
with the coefficients $c_{s}$ defined as
\begin{equation}
c_{s}=\frac{1}{\langle \Phi_{s}\bar{V}\Phi_{s}\rangle}\langle \Phi_{s}\bar
{V}\mathcal{F}_{1}^{(N)}\rangle,\  \ s=1,2,..,N.\label{43}%
\end{equation}
These coefficients are obtained algebraically by solving the matrix equation
(\ref{44}) for the $c_{s}$%
\begin{equation}
\sum_{s^{\prime}=1}^{N}(\delta_{s,s^{\prime}}-\mathcal{B}_{s.s^{\prime}%
})\ c_{s^{\prime}}=\frac{1}{\langle \Phi_{s}\bar{V}\Phi_{s}\rangle}\langle
\Phi_{s}\bar{V}F\rangle,\  \ s=1,2,..,N.\label{44}%
\end{equation}
with
\begin{equation}
\mathcal{B}_{s,s^{\prime}}=\frac{1}{\langle \Phi_{s}\bar{V}\Phi_{s}\rangle
}\langle \Phi_{s}\bar{V}\mathcal{O\ }\Phi_{s^{\prime}}\rangle \label{45}%
\end{equation}
This equation is obtained \cite{CANTON} by multiplying Eq. (\ref{42}) on the
left by $\Phi_{s}(r)\bar{V}(r)/\langle \Phi_{s}\bar{V}\Phi_{s}\rangle$ and
integrating over $r$ from $0$ to $\infty.$ The iterative solution of Eq.
(\ref{22})
\begin{equation}
\psi=\mathcal{F}_{1}^{(N)}+\chi^{(1)}+\chi^{(2)}+\chi^{(3)}+...\label{46}%
\end{equation}
is obtained by solving
\begin{equation}
\chi^{(n+1)}=\mathcal{O}_{N}\chi^{(n+1)}+\mathcal{O}_{R}\chi^{(n)}%
,\  \ n=1,2,...\label{47}%
\end{equation}
iteratively with $\chi^{(1)}=\mathcal{F}_{1}^{(N)}.$ This series will converge
if the norm of $\mathcal{O}_{R}$ is less than unity. The advantage of this
procedure is that the functions $\chi^{(n)}$ can also be obtained
algebraically, using the same matrix as required for the calculation of the
coefficients $c_{s}$, and the evaluation of $\mathcal{O}_{R}\chi^{(n)}$ only
involves integrals over known functions, rather than the solution of linear
equations, which would be more computer intensive. The matrices are of
dimension $N\times N$, where $N$ is of order $50$ or less. Numerical examples
are given below.

In the approximate Q-P method $\mathcal{S}_{\emph{2}}$ one solves the once
iterated form of Eq. (\ref{22a}) $\psi=F\ +\  \mathcal{O(}F\mathcal{+O}%
\psi \mathcal{)}$%
\begin{equation}
\psi=F\ +\mathcal{O}F\mathcal{+O}^{2}\psi.\label{48}%
\end{equation}
The iterative treatment is similar to that of method $\mathcal{S}_{1}$, with
the driving function $F$ replaced by $F\ +\mathcal{O}F$, and the operator
$\mathcal{O}$ in Eq. (\ref{22a}) replaced by $\mathcal{O}^{2}.$ This method
lends itself to a decomposition of the matrices of $\mathcal{O}^{2}$ in terms
of the singular value decomposition method (SVD) \cite{ZERRAD}, \cite{SVD},
that takes the place of finding the eigenvalues and eigenvectors $\gamma_{i}$
and $\Gamma_{i}$ of the operator $\mathcal{O}$, and provides a measure of the
norm of $(\mathcal{O}^{2})_{R}.$ This method is described in Ref.
\cite{RAWSVD}, but is not used for the numerical calculations described in the text.

\section{Numerical examples for methods $\mathcal{S}_{1}$ and $\mathcal{S}%
_{2}$}

The operator $\mathcal{O}(r,r^{\prime})$ for all of the three numerical
examples described below is $\mathcal{G}_{0}(r,r^{\prime})V(r^{\prime}),$
where $\mathcal{G}_{0}$ is the Green's function defined in Eqs. (\ref{4}) and
(\ref{5}), $V$ is the scattering potential\ given by $V_{P}$, defined in Eq.
(\ref{21}) and Table \ref{TABLE1}. The integral equation being solved is the
conventional Lippmann-Schwinger ($L-S$) scattering equation, Eq. (\ref{3}).
For most cases the wave number $k$ that defines the energy $E=k^{2}$ in Eq.
(\ref{2}) has the value $k=0.5\ fm^{-1}.$ As illustrated in Fig. \ref{FIG1},
$V_{P}$ has a repulsive core near the origin followed by an attractive well
that decreases exponentially with distance. The original $Q-P$ method is
illustrated by example $P,$ and the approximate Q-P method $\mathcal{S}_{2}$
is illustrated for two sturmian potentials $V_{S}$ and $V_{WS}.$ The size $N$
of the sturmian basis is $46$, $50$, and $41$ for potentials $V_{S}$, $V_{B}$,
and $V_{WS},$ respectively

\subsection{Q-P example "P"}

In this case the sturmian potential is equal to the scattering potential,
$\bar{V}=V_{P},$ and hence the sturmian functions $\Phi_{s}$, defined by Eq.
(\ref{10}) are the same, to within a normalization constant, as the
eigenfunctions $\Gamma_{i}^{(N)}$ of the operator $\mathcal{O=G}_{0}V_{P}$,
and the eigenvalues $\eta_{s}$ are also the same as the eigenvalues
$\gamma_{i}^{(N)}$\ The $\Phi_{s}$ used in the present examples are normalized
such that asymptotically they equal the function $H(r)$, with unit
coefficient. If an eigenfunction $\bar{\Phi}_{s}$ is normalized differently,
then it can be renormalized according to%
\begin{equation}
\Phi_{s}=\frac{-\eta_{s}k}{\langle FV\bar{\Phi}_{s}\rangle}\bar{\Phi}%
_{s}\label{33}%
\end{equation}
where $\langle FV\bar{\Phi}_{s}\rangle=\int_{0}^{r_{\max}}F(r)V(r)\  \bar{\Phi
}_{s}(r)\ dr.$ The value of $k$ is $0.5fm^{-1}$, and $r_{\max}=30\ fm$. Some
of these sturmians are illustrated in Figs. \ref{FIG6} to \ref{FIG9} and the
absolute values of $\eta_{s}$ are displayed in Fig. \ref{FIG11}$.$ The
operator $\mathcal{O}_{N}$ becomes
\begin{equation}
\mathcal{O}_{N}=\sum_{s=1}^{N}\Phi_{s}(r)\frac{\eta_{s}}{\langle \Phi_{s}%
V\Phi_{s}\rangle}\Phi_{s}(r^{\prime})V(r^{\prime}).\label{34}%
\end{equation}
In the present example $N=26$, and $|\eta_{26}|=0.1425$. The matrix elements
in Eq. (\ref{44}) become diagonal, the coefficients $c_{s}$, defined in Eq.
(\ref{44}), are given by $\langle FV\Phi_{s}\rangle/[(1-\eta_{s})\langle
\Phi_{s}V\Phi_{s}\rangle],$ and the function $\mathcal{F}^{(N)}$, solution of
Eq. (\ref{30}), is given by%
\begin{equation}
\mathcal{F}_{1}^{(N)}\mathcal{(}r\mathcal{)=\ }F(r)\  \mathcal{+}\sum_{s=1}%
^{N}\frac{\eta_{s}}{1-\eta_{s}}\frac{\langle FV\Phi_{s}\rangle}{\langle
\Phi_{s}V\Phi_{s}\rangle}\Phi_{s}(r),\label{36}%
\end{equation}
The quantity $S_{1}$ is defined by the asymptotic limit of $\mathcal{F}$
\begin{equation}
\mathcal{F}_{1}^{(N)}\mathcal{(}r\rightarrow \infty \mathcal{)=\ }%
F(r)+S_{1}\ H(r).\label{38}%
\end{equation}
and is given by $S_{1}=\sum_{1}^{M}\bar{c}_{s}$ , where
\begin{equation}
\bar{c}_{s}=\frac{\eta_{s}}{1-\eta_{s}}\frac{\langle FV\Phi_{s}\rangle
}{\langle \Phi_{s}V\Phi_{s}\rangle}.\label{37}%
\end{equation}
and the iterative corrections to $\psi$ are
\begin{equation}
\psi=\mathcal{F}_{1}^{(N)}\mathcal{+O}_{R}F\mathcal{+O}_{R}^{2}F\mathcal{+O}%
_{R}^{3}F\mathcal{+}\ ...,\label{39}%
\end{equation}
as shown in the Appendix A.

\begin{figure}
[ptb]
\begin{center}
\includegraphics[
height=1.7426in,
width=2.316in
]%
{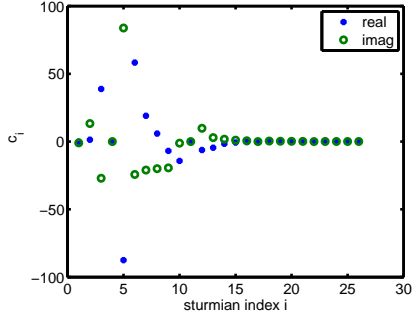}%
\caption{The coefficients $\bar{c}_{i}$ as defined in Eq. (\ref{37}) for the
Q-P method. The sturmian potential is the same as the scattering potential
$V_{P}$ and $k=0.5\ fm^{-1}.$}%
\label{FIG12}%
\end{center}
\end{figure}

\begin{figure}
[ptb]
\begin{center}
\includegraphics[
height=1.8784in,
width=2.4967in
]%
{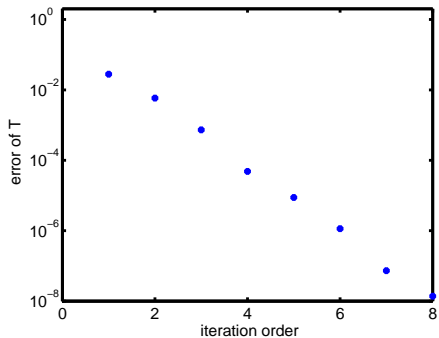}%
\caption{The iterative improvement of the asymptotic value of $S$ of the
scattering wave function $\psi$ is shown as the absolute value of the
difference between the iterative value of $S,$ obtained from Eq. (\ref{39})
and the value of $\psi$, obtained by solving the Lippmann-Schwinger scattering
equation directly. The first point illustrates the difference between the
asymptotic values of $\mathcal{F}$ and $\psi$, while the subsequent points
illustrate the contributions of the successive Weinberg iterations.}%
\label{FIG13}%
\end{center}
\end{figure}

The values of $\bar{c}_{i}$ are shown in Fig. \ref{FIG12}. They begin to
decrease nearly monotonically for $i>15,$ but the rate of decrease is slow.
For example $\bar{c}_{15}=-0.7017+1.0067i,$ and $\bar{c}_{26}%
=-0.0107+0.0395i.$ This shows that without the additional iterations described
next, the accuracy of $S$ given by $S_{1}$ alone$,$ would be good to only one
or two significant figures. For example, the value of $S_{1}$ obtained from
$\mathcal{F}^{(N)}$ is $S_{1}=0.7075982522+0.493097515i,$ while the "exact"
value of $S$, obtained via the spectral Integral Equation Method (IEM)
\cite{SPECTRAL-A} is $S=0.34016623828+0.8664518117i.$ In Fig. \ref{FIG13} the
improvement due to the successive Weinberg iterations are displayed.%

\subsection{Examples for the approximate Q-P Methods $\mathcal{S}_{1}$ and
$\mathcal{S}_{2}.$}

For both methods $\mathcal{S}_{1}$ and $\mathcal{S}_{2}$ the operator
$\mathcal{O}$ is again $\mathcal{G}_{0}V,$ with $V=V_{P}$, and the sturmian
potential $\bar{V}$ can be either $V_{S},$ Eqs. (\ref{21}), or $V_{B},$ the
latter defined in Appendix $B.$ In this section the normalization of the
sturmian functions is such that%
\begin{equation}
\langle \Phi_{s}\bar{V}\Phi_{s}\rangle=\eta_{s}~~,s=1,2,..,N \label{55}%
\end{equation}
For method $\mathcal{S}_{1}$ the elements of the matrix $\mathcal{B}$,
Eq.$\ (\ref{45}),$ become $\mathcal{B}_{s,s^{\prime}}=\langle \Phi_{s}\bar
{V}\Phi_{s^{\prime}}\rangle \eta_{s}/\langle \Phi_{s}\bar{V}\Phi_{s}\rangle$,
and all other equations described in section III apply.\ The separable
approximation to the operator $\mathcal{O},$ Eq. (\ref{25}), is given by
\begin{equation}
\mathcal{O}_{N}\  \mathcal{=}\sum_{s=1}^{N}\mathcal{G}_{0}V\Phi_{s}\rangle
\frac{1}{\langle \Phi_{s}\bar{V}\Phi_{s}\rangle}\langle \Phi_{s}\bar{V},
\label{56}%
\end{equation}
and as a result of the normalization (\ref{55}) the $n$'th power of
$\mathcal{O}_{N}\ $is%
\begin{equation}
\left(  \mathcal{O}_{N}\right)  ^{n}\  \mathcal{=}\sum_{s,s^{\prime}=1}%
^{N}\mathcal{G}_{0}V\Phi_{s}\rangle \left(  \boldsymbol{V}^{\ n-1}\right)
_{s,s^{\prime}}\frac{1}{\eta_{s^{\prime}}}\langle \Phi_{s^{\prime}}\bar{V},
\label{57}%
\end{equation}
where the matrix $\boldsymbol{V}$ has the matrix elements%
\begin{equation}
\boldsymbol{V}_{s,s^{\prime}}=\langle \Phi_{s}V\Phi_{s^{\prime}}\rangle.
\label{58}%
\end{equation}
Equation (\ref{57}) shows that if the norm of the matrix $\boldsymbol{V}$ is
larger than unity, iterations performed with $(\mathcal{O}_{N})^{2}\ $will not
converge. For this reason separating from $\boldsymbol{V}$ a part that has a
norm less than unity, and performing iterations for $(\mathcal{O}_{N})^{2}$ on
that part, will converge. That is the reason for solving the iterated equation
(\ref{48}) rather than the original Eq. (\ref{3}), as will be explained
further below. \ For more complicated integral kernels a similar separation is
also feasible with the Singular Value Decomposition method (SVD) \cite{RAWSVD}
but is not needed for the examples given here.

In method $\mathcal{S}_{2}$ \ one decomposes $\mathcal{O}^{2},$\ the square of
the full operator, into two parts $\mathcal{O}^{2}=\mathcal{O}_{N}%
^{2}+\mathcal{O}_{R}^{2}.$ According to Eq. (\ref{57}) ($\mathcal{O}_{N})^{2}$
is given by
\begin{equation}
\mathcal{O}_{N}^{2}=\sum_{s,s^{\prime}=1}^{N}\mathcal{G}_{0}V\Phi_{s}%
\rangle \boldsymbol{V}_{s,s^{\prime}}\frac{1}{\eta_{s^{\prime}}}\langle
\Phi_{s^{\prime}}\bar{V} \label{59}%
\end{equation}
and one $\mathcal{O}_{R}^{2}$\ is defined as
\begin{equation}
\mathcal{O}_{R}^{2}=\mathcal{O}^{2}-\mathcal{O}_{N}^{2} \label{60}%
\end{equation}

In order to solve the once iterated $L-S$ equation (\ref{48}), one defines the
function $\mathcal{F}_{2}$ as the solution of
\begin{equation}
\mathcal{F}_{2}^{(N)}=F\ +\mathcal{O}F\mathcal{+O}_{N}^{2}\mathcal{F}%
_{2}^{(N)} \label{62}%
\end{equation}
given in the present case by%
\begin{equation}
\mathcal{F}_{2}^{(N)}=F\ +\mathcal{O}F+\sum_{s,s^{\prime}=1}^{N}%
\mathcal{G}_{0}V\Phi_{s}\rangle \boldsymbol{V}_{s,s^{\prime}}c_{s^{\prime}%
}^{(2)} \label{63}%
\end{equation}
where $c_{s^{\prime}}^{(2)}=$\ $\langle \Phi_{s^{\prime}}\bar{V}\mathcal{F}%
_{2}^{(N)}\rangle/\eta_{s^{\prime}}$is the solution of the linear equation%
\begin{equation}
\sum_{s^{\prime}=1}^{N}(\delta_{s,s^{\prime}}-\boldsymbol{V}_{s,s^{\prime}%
}^{2})c_{s^{\prime}}^{(2)}=\frac{1}{\eta_{s}}\langle \Phi_{s}\bar
{V}(F+\mathcal{O}F)\rangle. \label{64}%
\end{equation}
The terms $\chi_{2}^{(n)}$ required for the subsequent iterations,%
\begin{equation}
\psi=\mathcal{F}_{2}^{(N)}+\chi_{2}^{(2)}+\chi_{2}^{(3)}+\chi_{2}^{(4)}+...
\label{65}%
\end{equation}
are obtained by solving
\begin{equation}
\chi_{2}^{(n+1)}=\mathcal{O}_{N}^{2}\chi_{2}^{(n+1)}+\mathcal{O}_{R}^{2}%
\chi_{2}^{(n)},\  \ n=1,2,... \label{66}%
\end{equation}
with $\chi^{(1)}=\mathcal{F}_{2}^{(N)}.$

A justification for the advantage of method $\mathcal{S}_{2}$ over
$\mathcal{S}_{1}$ is as follows. Formally, Eq. (\ref{66}) can be written as
$\chi_{2}^{(n+1)}=\left[  (1-\mathcal{O}_{N}^{2})^{-1}\mathcal{O}_{R}%
^{2}\right]  ^{n}\mathcal{F}_{2}^{(N)}$, and hence the rate of convergence of
the iterations depends on the norm of the operator $(1-\mathcal{O}_{N}%
^{2})^{-1}\mathcal{O}_{R}^{2}.$ Since the norm of $(1-\mathcal{O}_{N}%
^{2})^{-1}$\ is expected to be smaller than unity, the rate of convergence
should be faster than the powers of the norm of $\mathcal{O}_{R}^{2}.$ In view
of the completeness of the sturmian functions, $\mathcal{O}_{R}^{2}$ is also
given by%
\begin{equation}
\mathcal{O}_{R}^{2}=\sum_{s,s^{\prime}=N+1}^{\infty}\mathcal{G}_{0}V\Phi
_{s}\rangle \boldsymbol{V}_{s,s^{\prime}}\frac{1}{\eta_{s^{\prime}}}\langle
\Phi_{s^{\prime}}\bar{V}. \label{61}%
\end{equation}
By defining $\boldsymbol{\tilde{V}}$%
\begin{align}
\boldsymbol{\tilde{V}}_{s,s^{\prime}}  &  =\boldsymbol{V}_{s,s^{\prime}%
}\text{\  \  \ }s,s^{\prime}>N\nonumber \\
\boldsymbol{\tilde{V}}s,s^{\prime}  &  =0\text{ \  \  \  \  \ }s,s^{\prime}\leq N
\label{68}%
\end{align}
and keeping in mind that
\begin{equation}
\frac{1}{\eta_{s^{\prime}}}\langle \Phi_{s^{\prime}}\bar{V}\mathcal{G}_{0}%
V\Phi_{s}\rangle=\boldsymbol{V}_{s^{\prime},s} \label{68a}%
\end{equation}
one can show that the powers of $\mathcal{O}_{R}^{2}$ are given by%
\begin{equation}
\left(  \mathcal{O}_{R}^{2}\right)  ^{n}=\sum_{s,s^{\prime}=N+1}^{\infty
}\mathcal{G}_{0}V\Phi_{s}\rangle \left(  \boldsymbol{\tilde{V}}^{\ 2n-1}%
\right)  _{s,s^{\prime}}\frac{1}{\eta_{s^{\prime}}}\langle \Phi_{s^{\prime}%
}\bar{V}, \label{69}%
\end{equation}
and hence the norm of the iterations should decrease with the order of the
iteration $n$ as%
\begin{equation}
|\chi_{2}^{(n)}|<\mathfrak{k}\left \vert \boldsymbol{\tilde{V}}^{\ 2n-3}%
\right \vert ,\  \  \ n=1,2,.. \label{70}%
\end{equation}
where $\mathfrak{k}$ is some constant.

By contrast, for method $\mathcal{S}_{1}$
\begin{equation}
\left(  \mathcal{O}_{R}\right)  ^{n}=\sum_{s,s^{\prime}=N+1}^{\infty
}\mathcal{G}_{0}V\Phi_{s}\rangle \left(  \boldsymbol{\tilde{V}}^{\ n-1}\right)
_{s,s^{\prime}}\frac{1}{\eta_{s^{\prime}}}\langle \Phi_{s^{\prime}}\bar
{V},\label{71}%
\end{equation}
and by a reasoning similar to the one above, one expects%
\begin{equation}
|\chi^{(n)}|<\mathfrak{k}^{\prime}\left \vert \boldsymbol{\tilde{V}}%
^{\ n-1}\right \vert ,\  \  \ n=1,2,..\label{72}%
\end{equation}
These convergence estimates are supported by Figs. \ref{21} and \ref{24}.%
\begin{figure}
[ptb]
\begin{center}
\includegraphics[
height=2.0678in,
width=2.7492in
]%
{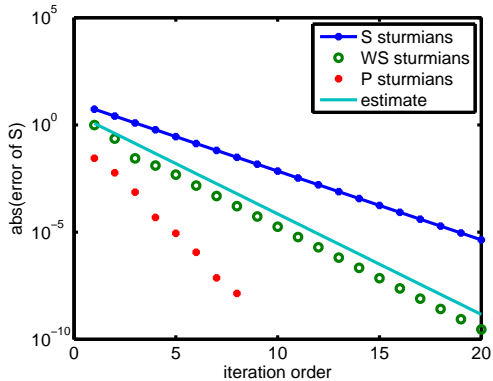}%
\caption{Iterative rate of convergence of the asymptotic limit of the wave
function to the "exact" one. The sturmian functions $S$ and $WS$ are obtained
with potentials $V_{S}$ and $V_{WS}$, respectively, which have no repulsive
core. Iteration method $\mathcal{S}_{2}$ was used for results $V_{S}$ and
$V_{WS}$ and method $\mathcal{S}_{1}$ was used for \ potential $V_{P}.$ (The
latter is identical to the potential used to calculate the scattering wave
function). The solid line represents $(0.66)^{p}$, with $p=2.6\times n-3,$ $n$
being the iteration order. The wave number is $k=0.5\ fm^{-1}.$}%
\label{FIG21}%
\end{center}
\end{figure}
Although one would expect from Eq. (\ref{70}) that the magnitude of the
iteration error decreases as the square of the iteration order $n$, (i.e. a
power $2$) the fit in Fig. \ref{21} decreases by a larger power, $2.6$.
Nevertheless, the convergence of method $\mathcal{S}_{1}$ is considerably
slower than for $\mathcal{S}_{2\text{ }},$ as illustrated in Fig. \ref{24},
and is in accordance with Eq. (\ref{72}).
\begin{figure}
[ptb]
\begin{center}
\includegraphics[
height=2.2399in,
width=2.9784in
]%
{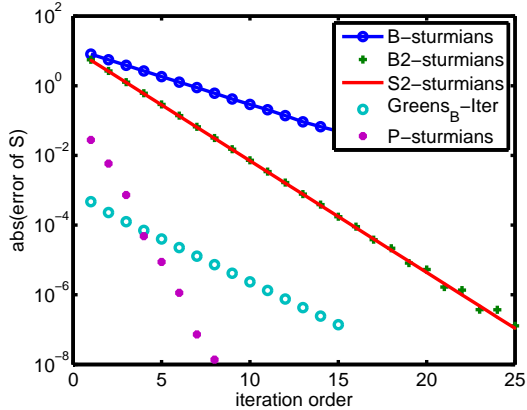}%
\caption{The convergence of the iterations $n=1,2,..$ as measured by the error
of the asymptotic value of the wave function $\psi,$ Eq. (\ref{6}). The
results labeled $B$ and $B2$ are obtained with sturmian potential $\bar
{V}=V_{B}$ for methods $\mathcal{S}$ and $\mathcal{S}_{2}$, respectively. The
solid line labeled as S2 is obtained with method $\mathcal{S}_{2}$ for the
sturmian potential $V_{S}.$ The result $P$ is obtained with the original Q-P
method, with sturmians obtained for potential $V_{P}.$ The open circles,
labeled $Green_{B},$ are obtained with Green's function iterations, based on
$V_{P}-V_{B}$, described in Appendix $B.$}%
\label{FIG24}%
\end{center}
\end{figure}

\begin{figure}
[ptb]
\begin{center}
\includegraphics[
height=2.2191in,
width=2.9507in
]%
{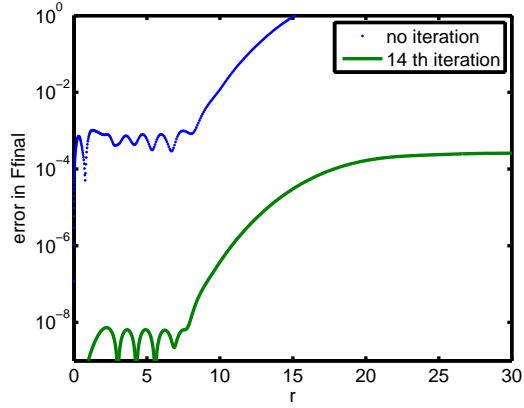}%
\caption{Absolute value of the error of the wave function $\psi$ as a function
of radial distance $r$ . The result labeled "no iteration" illustrates the
error of the function $\mathcal{F}_{2}^{(N)}$, Eq. (\ref{63}), with $N=46.$
The other line is obtained after the $14$'th iteration. These resuts are
obtained with method $\mathcal{S}_{2}$ using the sturmians for potential
$V_{S}$. The scattering wave function $\psi$ is obtained with potential
$V_{P}$ for $k=0.5\ fm^{-1}.$ }%
\label{FIG22}%
\end{center}
\end{figure}

The results for method $\mathcal{S}_{2}$, displayed in Fig. (\ref{FIG21}),
also show that the iterations using sturmians that are based on a potential
with a longer range, $V_{WS},$ than the potential $V_{S}$ converge faster.
This feature can be understood from the fact that at large distances the
sturmians become linearly dependent, since asymptotically they become
proportional to each other, as can be seen from Figs. \ref{FIG4} to
\ref{FIG7}, and hence at large distances they do not form a good basis set.
Indeed, at short distances the convergence to the scattering wave function is
much faster than at large distances, as can be seen from Fig. \ref{FIG22},
obtained with sturmians for $V_{S}.$

\begin{figure}
[ptb]
\begin{center}
\includegraphics[
height=2.1404in,
width=2.8461in
]%
{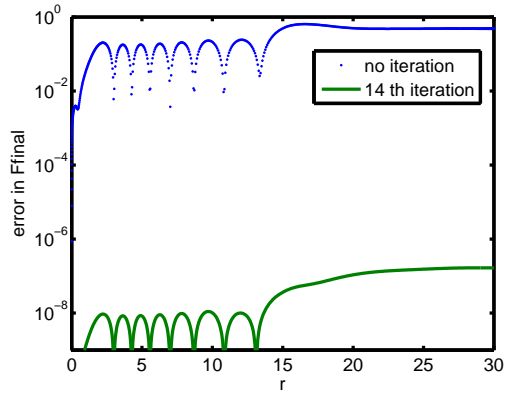}%
\caption{Same as Fig. \ref{FIG22}, using $41$ sturmians based on the $V_{WS}$
potential.}%
\label{FIG23}%
\end{center}
\end{figure}

The results labeled $B$ and $B2$ in Fig. \ref{FIG24} are obtained with methods
$S$ and $S_{2,}$ respectively, for sturmians based on potential $V_{B}.$ This
potential, defined in Eq. (\ref{21WS}) and further illustrated in the
Appendix, is identical to potential $V_{P}$ at large distances, but has its
repulsive core decreased by a factor that decreases to zero near the origin.
Both potentials $V_{B}$ and $V_{S}$ decay in a similar fashion at large
distances while $V_{B}$ has a barrier near the origin, and $V_{S}$ does not.
The result that both sets of sturmians give nearly indistinguishable results
for the iteration, as shown in Fig. \ref{FIG24}, shows that the behavior of
the sturmians near the origin does not significantly affect the results.
However, using the sturmians defined with the longer range potential $V_{WS}$
the convergence is significantly better at large distances, as is shown in
Fig. \ref{FIG23}.%

The open circles in Fig. \ref{24}, labeled as $Greens_{B}$ were obtained with
a Green's function iteration method, described in Appendix $B$, in which
potential $V_{p}$ is divided into $V_{B}+(V_{P}-V_{B}).$ The $L-S$ equation
with potential $V_{B}$ is solved to produce the function $\mathcal{F}$ and the
corrections due to $(V_{P}-V_{B})$ are obtained iteratively in a Born-series
manner as approximation to the exact function $\psi$. The asymptotic value of
$\mathcal{F}$ is much closer to that of $\psi$ than for method $\mathcal{S}%
_{2}$, but the rate of convergence of the Green's function iterations is not
as fast as that of method $\mathcal{S}_{2}.$

\section{Summary and Conclusions}

The main result of this paper is to present a viable iteration procedure to
correct for the truncation errors of a sturmian expansion used for the
solution of an integral equation. The basic idea is to construct a separable
(and truncated) approximation to the integral kernel of the integral equation,
that leads to an algebraic solution for the approximated integral equation,
denoted as $\mathcal{F}.$ The corrections to $\mathcal{F}$ are performed
iteratively, and the rate of convergence of the iteration is investigated. The
advantage of doing iterations rather than solving the integral equation
numerically directly, is that the numerical complexity for obtaining the
algebraic solution for $\mathcal{F}$ and the subsequent iterations can be
substantially less than for the direct numerical solution, especially if the
kernel of the integral equation is very complicated. An example would be the
solution of a Schr\"{o}dinger equation with nonlocal potentials, recast into a
Lippmann-Schwinger ($L-S)$ integral form. The numerical examples are done in
configuration space for a one-dimensional ($L-S)$ equation for which the
integral kernel is of the form $\mathcal{G}_{0}V,$ where $\mathcal{G}_{0}$ is
the undistorted Green's function and $V$ the scattering potential, and the
scattering energy is positive. However the formalism can be generalized to
more general integral kernels. Several methods are compared. One is the
"classical"\ quasi-particle \cite{WEINBERG} method, that requires a basis of
sturmians that are eigenfunctions of the kernel of the integral equation. For
the application of this method to the scattering of a particle from a
potential with a repulsive core, $V_{P},$ the iterations converge very fast
(denoted as method $P)$ and to high accuracy, as illustrated in Figs.
\ref{FIG13}, \ref{21}, and \ref{24}$.$ If the sturmian eigenstates of the
above mentioned integral kernel are not available, as is assumed here, then a
set of auxiliary sturmian functions are introduced, and the convergence of the
iterations is examined. Two methods are presented for this case. In method
$\mathcal{S}_{1}$ the iterative solution of the original integral equation is
examined, while in method $\mathcal{S}_{2}$ the solution of a \emph{once
iterated} integral equation is considered. It is predicted and also found that
the iterations of method $\mathcal{S}_{2}$ converge faster than for method
$\mathcal{S}_{1}$ , and both methods converge faster than the iterations
performed with a Green's function Born approximation procedure, but more
slowly than the original Q-P method. A key condition for obtaining an accurate
approximation to the wave function at large distances is that the auxiliary
sturmian functions be based on an auxiliary potential whose range is
sufficiently larger than that of the scattering potential. An alternate method
for solving the once iterated (but complicated) integral equation, based on
the singular value decomposition of a matrix, is described in Appendix $C$ but
not used in the numerical examples.

An accuracy of at least eight significant figures is obtained for all the
numerical calculations performed. This accuracy is achieved by using the
spectral method for solving integral equations \cite{SPECTRAL-B}, both for the
scattering wave function and for the sturmian functions. As a result, previous
investigations with positive energy sturmians \cite{RAW}, hampered by lack of
this type of accuracy, were improved upon and expanded in the present study.

\bigskip

{\LARGE Appendix A}

A proof of Eq. (\ref{39}) is as follows: \ The operator $\mathcal{O}_{N}$,
given by Eq. (\ref{34}), projects any function into the space spanned by the
sturmians $\Phi_{s}$, $s=1,2,..,N$, while the operator $\mathcal{O}_{R}$
projects any function into the sturmian space for $s=N+1,N+2,..$. The latter
property follows from the orthogonality given by Eq. (\ref{11}), and the fact
that the sturmians form a complete set. Hence $\mathcal{O}_{R},$
Eq.(\ref{27}), can also be written as
\begin{equation}
\mathcal{O}_{R}=\sum_{N+1}^{\infty}\Phi_{s}[\eta_{s}/\langle \Phi_{s}V\Phi
_{s}\rangle]\langle \Phi_{s}V\label{A15}%
\end{equation}
This expression for $\mathcal{O}_{R}$ also shows that the norm of
$\mathcal{O}_{R}$ is less than unity if
$\vert$%
$\eta_{s}|<1$ for $s>N.$ In view of Eqs. (\ref{30}), and (\ref{A15}) one finds
that $\mathcal{O}_{R}\mathcal{F}_{1}^{(N)}=\mathcal{O}_{R}F.$ By writing%
\begin{equation}
\psi=\mathcal{F}_{1}^{(N)}+\chi \label{A1}%
\end{equation}
and by by making use of Eqs. (\ref{47}) for $n=1$, $(1-$ $\mathcal{O}_{N}%
)\chi^{(1)}=\mathcal{O}_{R}\mathcal{F}^{(N)}=\mathcal{O}_{R}F.$ Hence
$\chi^{(1)}$ lies entirely in the sturmian space with $s>N$ and $\chi
^{(1)}=\mathcal{O}_{R}F.$ In view of Eq. (\ref{47}) the same argument can be
made for $\chi^{(2)},$ $\chi^{(3)}..$ and Eq. (\ref{39}) is proven for the
case that $\mathcal{O}=G_{0}V$ and $\bar{V}=V$.

\bigskip

{\LARGE Appendix B:\ The Green's function iterations.}

\begin{figure}
[ptb]
\begin{center}
\includegraphics[
height=1.535in,
width=2.041in
]%
{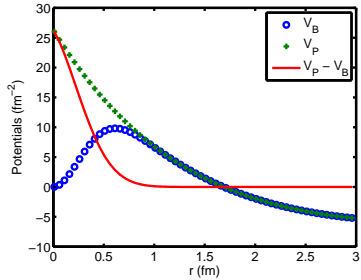}%
\caption{The potental $V_{B}\ $is illustrated by the open circles, and is
defined in Eq. (\ref{67}). The purpose of this potential is to eliminate the
repulsive core of potential $V_{P}$, the latter illustrated by the $+$
symbols. The difference between these two potentials, illustrated by the solid
line, serves to perform iterations with a Green's function Born-type method.}%
\label{FIG25}%
\end{center}
\end{figure}

At large distances potential $V_{B}$ is identical to $V_{P}$, but at short
distances the repulsive core is reduced by the factor $f(r)=1-\exp
[(r/0.5)^{2}]$, as given by Eq. (\ref{21WS})%
\begin{equation}
V_{B}(r)=f(r)\times V_{P}(r) \label{67}%
\end{equation}
This potential is illustrated by the open circles in Fig. \ref{FIG25}%
This potential was defined for the purpose of performing a Green's function
iteration, as follows. The potential $V_{P}$ is divided into two parts $V_{B}$
and $\ V_{2}=(V_{P}-V_{B})$. The first part has a suppressed repulsive core,
and the second part, $V_{2}$ adds the repulsive core again. The iterations are
based on Green's functions that explicitly include the distortion due to
$V_{B}$, and iterate over the effect of $V_{2}$ in a fashion similar to the
Born approximation. The Green's function method has many applications, but in
view of the factor $1/k$ in the Greens functions $F$ and $H,$ Eq. (\ref{4}),
this method does not converge well \cite{ZERRAD} for low values of the wave
number $k.$\bigskip

{\LARGE Appendix C:\ Iterative method to calculate Sturmian eigenfunctions.}

In what follows the eigenvalue subscripts $s$ will be dropped, and the
iterations which lead to the solution for a particular fixed value of $s$ will
be described. In the differential equation (\ref{8}) the exact values of
$\Lambda$ and $\Phi$ are initially unknown, but the equation can still be
solved for a guessed value $\Lambda^{(1)}$. One solutions, denoted as $Y(r)$
satisfies the correct boundary conditions at the origin, and is integrated
from inside outward.%
\begin{equation}
\mathcal{D}\ Y=\Lambda^{(1)}\bar{V}\ Y,\  \ 0\leq r\leq R_{I} \label{12}%
\end{equation}
The other solution, denoted as $Z(r)$, satisfies the correct boundary
conditions asymptotically,
\begin{equation}
\mathcal{D}\ Z=\Lambda^{(1)}\bar{V}\ Z,\  \ R_{I}\leq r\leq \infty, \label{13}%
\end{equation}
and is integrated inward towards the intermediate point $R_{I}$ chosen to lie
within the region where $\bar{V}$ is non negligible. The three solutions at
the point $R_{I}$ are denoted as $\Phi_{I},Y_{I}$,and $Z_{I}.$ One can
renormalize $Y$ by a factor $\mathfrak{k=}Z_{I}/Y_{I}$
\begin{equation}
\bar{Y}(r)=\mathfrak{k}\ Y(r) \label{14}%
\end{equation}
so that $\bar{Y}_{I}=Z_{I}.$

If one multiplies Eq. (\ref{8}) on both sides by $\bar{Y},$ and Eq. (\ref{12})
by $\Phi$, integrates over $r$ from $0$ to $R_{I}$, and subtracts the results
one from the other, after an integration by parts one obtains%
\begin{equation}
(\bar{Y}\  \Phi^{\prime}-\bar{Y}^{\prime}\  \Phi)\ |_{0}^{R_{I}}=(\Lambda
-\Lambda^{(1)})\  \int_{0}^{R_{I}}\bar{Y}(r)\  \bar{V}(r)\  \Phi(r)\ dr,
\label{15}%
\end{equation}
where the prime denotes a derivative with respect to $r.$ By a similar
procedure one obtains%
\begin{equation}
(Z\  \Phi^{\prime}-Z^{\prime}\  \Phi)\ |_{R_{I}}^{\infty}=(\Lambda-\Lambda
^{(1)})\  \int_{R_{I}}^{\infty}Z(r)\  \bar{V}(r)\  \Phi(r)\ dr. \label{16}%
\end{equation}
By adding Eqs. (\ref{15}) and (\ref{16}), by noting that $(\bar{Y}%
\  \Phi^{\prime}-\bar{Y}^{\prime}\  \Phi)\ |_{0}=0$ because $\bar{Y}$ and $\Phi$
vanish at the origin, and that $(Z\  \Phi^{\prime}-Z^{\prime}\  \Phi
)\ |^{\infty}=0$ because $Z$ and $\Phi$ obey the same boundary condition
asymptotically (to within a normalization constant), one obtains%
\begin{equation}
\Phi_{I}(Z_{I}^{\prime}-\bar{Y}_{I}^{\prime})=(\Lambda-\Lambda^{(1)})\  \left[
\int_{0}^{R_{I}}\bar{Y}(r)\  \bar{V}(r)\  \Phi(r)\ dr+\int_{R_{I}}^{\infty
}Z(r)\  \bar{V}(r)\  \Phi(r)\ dr\right]  . \label{17}%
\end{equation}
The above equation is still rigorously valid. The approximation consists in
replacing $\Phi(r)$ in the first integral by $\bar{Y}(r),$ and by $Z(r)$ in
the second integral, and further, by replacing $\Phi_{I}$ in the left hand
side by either $\bar{Y}_{I}$ or $Z_{I}.$ After making these approximations,
and by dividing both sides of Eq. (\ref{17}) by $(\bar{Y}_{I})^{2}=(Z_{I}%
)^{2},$ one obtains the final result%
\begin{equation}
\Lambda^{(2)}=\Lambda^{(1)}+\frac{Z_{I}^{\  \prime}/Z_{I}-Y_{I}^{\  \prime
}/Y_{I}}{\left(  \int_{0}^{R_{I}}Y^{\ 2}(r)\  \bar{V}(r)\ dr\right)
/Y_{I}^{\ 2}+\left(  \int_{R_{I}}^{\infty}Z^{\ 2}(r)\  \bar{V}(r)\ dr\right)
/Z_{I}^{\ 2}}\  \label{18}%
\end{equation}
In the above, $\Lambda^{(2)}$ replaces $\Lambda$ as the iteratively corrected
value for $\Lambda^{(1)}$; also the factor $\mathfrak{k}$ disappeared because
it canceled in the numerator and denominator of Eq. (\ref{18}).

Equation (\ref{18}) is the generalization to sturmian eigenvalues of the
iterative calculation of energy eigenvalues, Eq. (7) of Ref. \cite{HEHE}. It
can also be applied to negative energies, and can be generalized to the cases
of coupling potentials or of non-local potentials, which however goes beyond
the scope of the present study.

The functions $Y$ and $Z$ can be calculated as the solutions of the
differential Eqs. (\ref{12}) and (\ref{13}) by any convenient finite
difference method, or they can be obtained as the solutions of the integral
equations%
\begin{equation}
Y(r)=F(r)+\Lambda^{(1)}\int_{0}^{R_{I}}\mathcal{G}_{0}(r,r^{\prime})\bar
{V}(r^{\prime})Y(r^{\prime})dr^{\prime} \label{19}%
\end{equation}
and%
\begin{equation}
Z(r)=H(r)+\Lambda^{(1)}\int_{R_{I}}^{\infty}\mathcal{G}_{0}(r,r^{\prime}%
)\bar{V}(r^{\prime})Z(r^{\prime})dr^{\prime} \label{20}%
\end{equation}
or else the whole iteration procedure can be bypassed by obtaining the
eigenvalues $\eta_{s}$ and eigenfunctions $\Phi_{s}$ of the integral operator
$\mathcal{G}_{0}\  \bar{V}$ contained in Eq. (\ref{10}). In the numerical
calculations described below, this latter procedure is carried out (with
questionable accuracy according to Ref \cite{I-STRING}) so as to provide the
initial values $\Lambda_{s}^{(1)}$ for the iteration, Eq. (\ref{18}). The
present implementation obtains the solution of Eqs. (\ref{19}) and (\ref{20})
by using the spectral Chebyshev expansion method, \cite{SPECTRAL-A},
\cite{SPECTRAL-B} because its high accuracy can be predetermined by the
specification of an accuracy parameter $tol$, and the method is by now well
tested. \bigskip

{\LARGE Appendix D}

For future benchmark purposes, the eigenvalues of Eq. (\ref{8}) are given in
the Table \ref{TABLE2 copy(1)}. The potential is $V_{S}$ defined in Eq.
(\ref{21}) and Table \ref{TABLE1}, the wave number is $k=0.5fm^{-1}.$%

\begin{table}[tbp] \centering
\begin{tabular}
[c]{|l||l|l|}\hline
& Real part of $\Lambda_{s}$ & Imag. part of $\Lambda_{s}$\\ \hline \cline{1-1}%
$1$ & $-0.03297806784$ & $-0.05633093256$\\ \hline
$2$ & $0.04181033607$ & $-0.12298817955$\\ \hline
$3$ & $-0.02140651197$ & $-0.23641901361$\\ \hline
$4$ & $0.21055103262$ & $-0.15935376881$\\ \hline
$5$ & $0.43743611684$ & $-0.17810671020$\\ \hline
$6$ & $0.71976101832$ & $-0.19289033454$\\ \hline
$7$ & $1.05649701588$ & $-0.20516494913$\\ \hline
$8$ & $1.44649394898$ & $-0.21538251834$\\ \hline
$9$ & $1.88887575723$ & $-0.22390738772$\\ \hline
$10$ & $2.38303461225$ & $-0.23106484062$\\ \hline
$11$ & $2.92855765503$ & $-0.23712639599$\\ \hline
$12$ & $3.52516147256$ & $-0.24230800967$\\ \hline
$13$ & $4.17264661513$ & $-0.24677772914$\\ \hline
$14$ & $4.87086853262$ & $-0.25066566005$\\ \hline
$15$ & $5.61971940360$ & $-0.25407296058$\\ \hline
$16$ & $6.41911669954$ & $-0.25707898962$\\ \hline
$17$ & $7.26899583793$ & $-0.25974669404$\\ \hline
$18$ & $8.16930533353$ & $-0.26212658464$\\ \hline
$19$ & $9.12000350403$ & $-0.26425965551$\\ \hline
$20$ & $10.1210561674$ & $-0.26617953514$\\ \hline
$21$ & $11.1724349888$ & $-0.26791408364$\\ \hline
$22$ & $12.2741162656$ & $-0.26948659119$\\ \hline
\end{tabular}
\caption{Eigenvalues  for Sturmian potential $V_{S}$ for $k=0.5 fm^{-1}$.}\label{TABLE2 copy(1)}%
\end{table}%

\bigskip
\end{document}